\documentclass[11pt,a4paper]{article}
\pdfoutput=1
\usepackage{jheppub}
\usepackage{cite}

\usepackage{multirow, graphicx,amssymb,url,mathrsfs,amsmath}
\usepackage{wrapfig,boxedminipage,subfigure,epsfig}
\usepackage{amsxtra,amstext,latexsym,dsfont,amsfonts}
\usepackage{color}
\usepackage[dvipsnames]{xcolor}
\usepackage{float}
\usepackage{slashed}
\usepackage{calligra}
\usepackage{enumerate}
\DeclareFontShape{T1}{calligra}{m}{n}{<->s*[2.2]callig15}{}
\DeclareMathAlphabet{\mathcalligra}{T1}{calligra}{m}{n}

\newcommand{\be}{\begin{equation}}
\newcommand{\ee}{\end{equation}}
\newcommand{\bea}{\begin{eqnarray}}
\newcommand{\eea}{\end{eqnarray}}

\title{Scaling exponents of Mellin amplitudes for deriving bounds on flat space S-matrices from bounds on chaos}

\author[]{Mitsuhiro Nishida}

\affiliation[]{School of Physics and Chemistry, Gwangju Institute of Science and Technology, 123 Cheomdan-gwagiro, Gwangju 61005, Korea}

\emailAdd{mnishida@gist.ac.kr}

\abstract{We study an inequality between a scaling exponent $A$ in the Regge limit of tree-level flat space S-matrices with external massless scalars and another scaling exponent $A'$ in the Regge limit of the corresponding four-point scalar conformal correlators by using scaling exponents of Mellin amplitudes. We derive $A'\ge A$, which leads to the Regge growth bound of tree-level flat space S-matrices from the chaos bound in the flat space limit of the AdS/CFT correspondence, from polynomial boundedness of the Mellin amplitudes for local bulk descriptions. We also show $A'=A$ from the conformal block expansion in the $t$-channel with finite intermediate spins when coefficients are not small in the flat space limit.
 }

\begin{document}
\maketitle



\section{Introduction}
In perturbative quantum field theories, flat space S-matrices can in principle be computed from given Lagrangians. If any constraints on the Lagrangians are not imposed, one can build infinitely many phenomenological models. To reduce the number of models to be considered, physicists constrain the flat space S-matrices from physical assumptions such as unitarity, analyticity, causality, and crossing symmetry.

The authors of \cite{Chowdhury:2019kaq} constrained the local Lagrangians and the four-point polynomial S-matrices with scalars, photons, and gravitons by using the following Classical Regge Growth (CRG) conjecture:
\begin{quote}
The flat space S-matrices of consistent classical theories never grow faster than $S^2$ with fixed $T$ at physical Mandelstam variables $S$ and $T$ for any possible normalized polarization vectors.
\end{quote}
See also \cite{Chakraborty:2020rxf} for classification of the three-point S-matrices, which are related to the four-point S-matrices for exchange diagrams. For external scalars, the CRG conjecture was proved using twice-subtracted dispersion relations of the non-perturbative S-matrices \cite{Haring:2022cyf}.

In the flat space limit of the AdS/CFT correspondence, the CRG conjecture is closely connected to the chaos bound of out-of-time-order correlators (OTOCs) in large $N$ CFTs \cite{Maldacena:2015waa}. The late-time behaviors of the four-point conformal OTOCs in Rindler coordinates are given by the Regge limit of conformal correlators in Minkowski spacetime. The AdS/CFT correspondence gives holographic expressions of these correlators by scattering in Rindler-AdS coordinates \cite{Cornalba:2006xk, Cornalba:2006xm, Cornalba:2007zb, Cornalba:2007fs, Cornalba:2008qf, Shenker:2013pqa,  Roberts:2014ifa, Shenker:2014cwa, Perlmutter:2016pkf, Ahn:2019rnq}, and the flat space limit of the scattering amplitudes in AdS spacetime gives the flat space S-matrices.

There is a subtlety in this connection about the order of taking the flat space limit. Specifically, the chaos bound is the constraint in the Regge limit of conformal correlators, and the CRG conjecture is the constraint in the bulk point limit \cite{Gary:2009ae, Heemskerk:2009pn, Okuda:2010ym, Penedones:2010ue, Maldacena:2015iua}. The authors of \cite{Chandorkar:2021viw} carefully  investigated the two limits and derived an inequality $A'\ge A$ for tree-level contact diagrams with external scalars, photons, and gravitons. Here, $A'$ is a scaling exponent for the chaos bound, and $A$ is a scaling exponent for the CRG conjecture. If the chaos bound  is valid for external spinning fields, the inequality $A'\ge A$ leads to the CRG conjecture $2\ge A$ from the chaos bound $2\ge A'$.

In order to fully understand the inequality $A'\ge A$, there are more things to check as commented in \cite{Chandorkar:2021viw}. One of them is the inequality for tree-level exchange diagrams, which is significant for a constraint on the spin of exchange fields. For example, the saturation of the chaos bound $A'=2$ can be caused by the graviton exchange in the $t$-channel \cite{Maldacena:2015waa}. Another one is in what cases an equality $A'=A$ holds.

In this paper, we study the above questions about $A'\ge A$ for tree-level bulk diagrams with four external massless scalar fields by using Mellin amplitudes \cite{Mack:2009mi}. From the viewpoint of the Mellin amplitudes, the difference between $A'$ and $A$ can arise from two different regions in the coordinate space of Mellin variables for the Regge limit of conformal correlators \cite{Costa:2012cb} and for the bulk point limit of the correlators \cite{Penedones:2010ue}. For tree-level contact diagrams and tree-level exchange diagrams in the $t$-channel, we focus on the four-point correlators of single-trace scalars that admit the conformal block expansion in the $t$-channel with finite intermediate spins. We derive the inequality $A'\ge A$ by using polynomial boundedness of the Mellin amplitudes, which is a criterion for effective descriptions by local bulk theories \cite{Fitzpatrick:2012cg}. By using a formula of bulk S-matrices with the Mellin representations of the conformal block expansion, we also show that the equality $A'=A$ holds when coefficients are not small in the flat space limit.

 Our approach can be regarded as a simplified version of the method that gives bounds on the Mellin representations in \cite{Caron-Huot:2021enk}. For sharp bounds on local bulk effective theories from the conformal bootstrap, there are technical difficulties overcome in \cite{Caron-Huot:2021enk}. Our analysis to derive only $A'\ge A$ is easy and can be applied to the conformal block expansion with the exchange of light single-trace operators.
 The Mellin amplitudes in unitary CFTs are meromorphic without branch cuts \cite{Mack:2009mi}, and one does not need to consider multiple sheets in the approach by the Mellin amplitudes.

This paper is organized as follows. In Section \ref{review}, a derivation of $A'\ge A$ for tree-level contact diagrams  by \cite{Chandorkar:2021viw} is reviewed. We introduce $A'$ and $A$ in terms of the Mellin amplitudes in Section \ref{mellin}. The inequality $A'\ge A$ is derived from polynomial boundedness of the Mellin amplitudes in Section \ref{pbm}. In Section \ref{icbe}, we also derive the equality $A'=A$ from the conformal block expansion with coefficients that are not small in the flat space limit. Section \ref{summary} is devoted to summary and discussion.

\section{Review: a derivation of \texorpdfstring{$A'\ge A$}{} for tree-level contact diagrams}\label{review}
We briefly review a derivation of the inequality $A'\ge A$ between scaling exponents of normalized four-point correlators for bulk tree-level contact diagrams in two different limits by \cite{Chandorkar:2021viw}. In this derivation, it is crucial that the correlators can be expressed by double power series expansions.

Let us consider $d$-dimensional conformal correlation functions of four operators inserted at
\begin{align}
\begin{aligned}
P_1=&( \cos\tau, \sin\tau, 1, 0, \vec{0}), \;\;\;\;\;\;\;\;\;\;\;\; P_3=( \cos\tau, \sin\tau, -1, 0, \vec{0}),\\
P_2=&( -1, 0, -\cos\theta, -\sin\theta, \vec{0}),
\;\;\; P_4=( -1, 0, \cos\theta, \sin\theta, \vec{0}),
\end{aligned}\label{cr}
\end{align}
where we use $(d+2)$-dimensional coordinates of an embedding space $\mathbb{R}^{0,2}\otimes \mathbb{R}^{d,0}$. Here, the first two coordinates are time-like, and the others are space-like. Cross ratios in CFTs with (\ref{cr}) are given by \cite{Chandorkar:2021viw}
\begin{align}
u:=&\frac{(P_2\cdot P_1)(P_3\cdot P_4)}{(P_2\cdot P_4)(P_3\cdot P_1)}=\frac{(\cos\tau-\cos\theta)^2}{4},\\
v:=&\frac{(P_4\cdot P_1)(P_2\cdot P_3)}{(P_2\cdot P_4)(P_3\cdot P_1)}=\frac{(\cos\tau+\cos\theta)^2}{4},\\
\sigma^2:=&u=\frac{(\cos\tau-\cos\theta)^2}{4}, \;\;\; \cosh^2\rho:=\frac{(1+u-v)^2}{4u}=\frac{(1-\cos\theta\cos\tau)^2}{(\cos\theta-\cos\tau)^2}.
\end{align}

Generally, Lorentzian correlation functions have singularities and brunch cuts, which lead to distinct sheets in the complex cross ratio space. These sheets are associated with causal configurations of the four points $P_i$. We focus on the following two sheets:
\begin{align}
&\pi-\theta>\tau>\theta \;\;\;\;\;\; \text{(Causally Regge)},\\
&\tau<\theta \;\;\;\;\;\;\;\;\;\;\;\;\;\;\;\;\;\;\; \text{(Causally Scattering)}.
\end{align}
We also consider two limits of the correlators:
\begin{align}
&\tau\to0, \;\;\; \theta\to0, \;\;\; a:=\frac{\tau}{\theta} \; \text{fixed} \;\;\; \text{(Regge limit)},\label{reggel}\\
&\tau\to0, \;\;\; \theta \; \text{fixed} \;\;\; \;\;\;\;\;\;\;\;\;\;\;\;\;\;\;\;\;\;\;\;\;\;\text{(small $\tau$ limit)}.\label{stl}
\end{align}
The cross ratio $\sigma$ with $a\to0$ in the Regge limit and with $\theta\to0$ in the small $\tau$ limit can be estimated by
\begin{align}
\sigma\sim\frac{\theta^2}{4}.\label{rst}
\end{align}

\subsection{Scaling behavior in the Regge limit}
Our targets are normalized four-point functions $G^{\text{norm}}$ defined by
\begin{align}
G^{\text{norm}}:=\frac{G}{G_{12}G_{34}},\label{gnff}
\end{align}
where $G_{12}$ and $G_{34}$ are two-point functions, and $G$ is an unnormalized four-point functions. 
Due to bulk-to-boundary propagators, amplitudes of bulk four-point Witten diagrams in AdS satisfy a transformation rule of $G$ under conformal transformations \cite{Witten:1998qj}.
The normalized functions $G^{\text{norm}}$ in the Regge limit for bulk tree-level contact diagrams scale as \cite{Chandorkar:2021viw}
\begin{align}
G^{\text{norm}} \sim \frac{H(a)}{\theta^{2A'-2}} \;\;\; \text{(Regge limit)}.\label{rlnc1}
\end{align}
Demanding analytic properties of the function $H(a)$, the value of $A'$ does not depend on whether the sheet is Causally Regge or Causally Scattering.

\subsection{Scaling behavior in the small \texorpdfstring{$\tau$}{} limit}
The authors of \cite{Chandorkar:2021viw} also studied scaling behaviors of four-point functions in the small $\tau$ limit given by bulk tree-level contact diagrams with scalars, photons, and gravitons. The scaling behaviors with respect to $\theta$ in the small $\tau$ limit are given by
\begin{align}
G^{\text{norm}}\propto \frac{1}{\theta^{2A-2}} \;\;\; (\theta\to0 \;\; \text{in the small $\tau$ limit} ),
\end{align}
where these singular behaviors are known as the bulk point singularity \cite{Maldacena:2015iua}, and this limit is called the bulk point limit.
Scaling behaviors of flat space S-matrices $\mathcal{T}(S,T)$ in the Regge limit $(S\to\infty, \; T \; \text{fixed})$ for the corresponding bulk tree-level contact diagrams in the flat space limit are
\begin{align}
\mathcal{T}(S,T)\propto S^{A} \;\;\; (S\to\infty, \;  \text{fixed} \; T), \label{sbfsm}
\end{align}
where $S$ and $T$ are Mandelstam variables.

\subsection{Comparison between the scaling behaviors in the two limits}\label{subsec:comp}
The correlation functions $G$ studied in \cite{Chandorkar:2021viw} have the following scaling behaviors:
\begin{align}
G\sim\frac{1}{\theta^{2\tilde{B}-2}a^{\Delta_1+\Delta_2+\Delta_3+\Delta_4+r-3}}H(\theta,a) \;\;\; (\text{$\theta$ and $a$ are small.}),\label{fccd}
\end{align} 
where $\Delta_i$ are scaling dimensions of the operators, and $r$ is the number of derivatives in the bulk contact interaction. The function $H(\theta,a)$ is nonzero at $a\to0$ for some $\theta$ and admits a double power series expansion with respect to $\theta$ and $a$. In the small $\tau$ limit, $a$ goes to zero, and $H(\theta,a)|_{a\to0}$ determines the scaling behavior with respect to $\theta$.

From the property of $H(\theta,a)$, one can derive the inequality between $A'$ and $A$. Let $n_{Regge}$ be the smallest power of $\theta$ in $H(\theta,a)$ and let $n_{BP}$ be the smallest power of $\theta$ in $H(\theta,a)|_{a\to0}$, where the subscript $BP$ means the bulk point limit. Since the two-point functions $G_{12}$ and $G_{34}$ scale as
\begin{align}
G_{12}\propto \frac{1}{\theta^{2\Delta_1}} \;\;\;(\Delta_1=\Delta_2), \;\;\; G_{34}\propto \frac{1}{\theta^{2\Delta_3}} \;\;\;(\Delta_3=\Delta_4),
\end{align}
$A'$ and $A$ are given by
\begin{align}
A'=\tilde{B}-\Delta_1-\Delta_3-n_{Regge}, \;\;\; A=\tilde{B}-\Delta_1-\Delta_3-n_{BP}.
\end{align}
By definition, $n_{Regge}\le n_{BP}$ holds, and we obtain the desired inequality $A'\ge A$.

\section{Scaling behaviors of Mellin amplitudes in two limits}\label{mellin}
Conformal correlation functions can be expressed in integral representations using Mellin amplitudes \cite{Mack:2009mi}. We describe how $A'$ and $A$ do not have to be the same in terms of scaling exponents of the Mellin amplitudes. We consider the Mellin amplitudes with four external single-trace scalar operators for bulk massless scalars. Our convention of the Mellin amplitudes follows the convention in \cite{Costa:2012cb}.

\subsection{Review: Mellin amplitudes}
We define the Mellin amplitudes $M(\delta_{ij})$ for connected parts of scalar four-point functions $G$ by \cite{Mack:2009mi, Costa:2012cb}
\begin{align}
G=\int[d \delta]M(\delta_{ij})\prod_{1\le i<j\le4}\Gamma(\delta_{ij})(-2P_i\cdot P_j)^{-\delta_{ij}},\label{mr}
\end{align}
where $\Gamma(x)$ is the gamma function, and the integration contours are parallel to the imaginary axis. The four-point functions (\ref{mr}) include six variables $\delta_{ij}=\delta_{ji}$ $(i\ne j)$ with four constraints $\sum_{j\ne i}\delta_{ij}=\Delta_i$, and therefore there are two independent variables. We define two Mandelstam-like variables as
\begin{align}
t:=\Delta_1+\Delta_2-2\delta_{12}, \;\;\; s:=\Delta_{34}-2\delta_{13}, \;\;\; \Delta_{ij}:=\Delta_i-\Delta_j.
\end{align}
For the scalar four-point functions $G$, we define normalized scalar four-point functions $G^{\text{norm}}(u,v)$ as
\begin{align}
G=&:\frac{1}{(-2P_1\cdot P_2)^{\frac{\Delta_1+\Delta_2}{2}}(-2P_3\cdot P_4)^{\frac{\Delta_3+\Delta_4}{2}}}\left(\frac{P_2\cdot P_4}{P_1\cdot P_4}\right)^{\frac{\Delta_{12}}{2}}\left(\frac{P_1\cdot P_4}{P_1\cdot P_3}\right)^{\frac{\Delta_{34}}{2}}G^{\text{norm}}(u,v),
\end{align}
where this definition agrees with (\ref{gnff}) when $\Delta_{12}=\Delta_{34}=0$ holds.
By using $[d \delta]=dt ds/(4\pi i)^2$ and $M(\delta_{ij})=M(s,t)$, we obtain the following Mellin representations of the normalized functions $G^{\text{norm}}(u,v)$ \cite{Costa:2012cb}:
\begin{align}
\begin{aligned}
G^{\text{norm}}(u,v)=&\int_{-i\infty}^{i\infty}\frac{dt ds}{(4\pi i)^2}M(s,t)u^{t/2}v^{-(s+t)/2}\Gamma\left(\frac{\Delta_1+\Delta_2-t}{2}\right)\Gamma\left(\frac{\Delta_3+\Delta_4-t}{2}\right)\\
\times&\Gamma\left(\frac{\Delta_{34}-s}{2}\right)\Gamma\left(\frac{-\Delta_{12}-s}{2}\right)\Gamma\left(\frac{t+s}{2}\right)\Gamma\left(\frac{t+s+\Delta_{12}-\Delta_{34}}{2}\right).
\end{aligned}\label{nma}
\end{align}
The integration contours should be chosen such that all poles of $M(s,t)$ and the gamma functions in (\ref{nma}) are located on one side of each complex plane divided by the contours.

The Regge limit of $G^{\text{norm}}(u,v)$ on the Causally Regge sheet can be calculated by an analytic continuation of $G^{\text{norm}}(u,v)$ on the Causally Euclidean sheet \cite{Cornalba:2006xk, Cornalba:2006xm, Cornalba:2007zb, Cornalba:2007fs, Cornalba:2008qf}. As derived in the appendix of \cite{Costa:2012cb}, the Regge limit on the Causally Regge sheet is controlled by the large $s$ behavior\footnote{More precisely, the behavior at $s=ix, x\to\infty$ is dominant.} of $M(s,t)$ with fixed $t$, and the Regge limit of the Mellin amplitude is defined by this large $s$ limit with fixed $t$. In particular, if the behavior of $M(s,t)$ in the Regge limit is
\begin{align}
M(s,t)\propto s^{A'} \;\;\; (s\to\infty,  \; \text{fixed} \; t), \label{rlma}
\end{align}
the Regge limit of $G^{\text{norm}}(u,v)$ on the Causally Regge sheet scales as
\begin{align}
G^{\text{norm}}(u,v) \propto \frac{1}{\sigma^{A'-1}} \;\;\; \text{(Regge limit on the Causally Regge sheet)},\label{rlnc2}
\end{align}
where we choose (\ref{rlma}) so that (\ref{rlnc2}) agrees with (\ref{rlnc1}) by using (\ref{rst}).

If we suppose the AdS/CFT correspondence in the flat space limit, there is a formula between the Mellin amplitudes $M(s,t)$ and the flat space S-matrices $\mathcal{T}(S,T)$ \cite{Penedones:2010ue, Fitzpatrick:2011hu}:
\begin{align}
\mathcal{T}(S,T)=&\frac{1}{\mathcal{N}}\lim_{R\to\infty}R^{2h-3}\int^{i\infty}_{-i\infty}\frac{d\alpha}{2\pi i} \alpha^{h-\sum_{i}\Delta_i/2}e^{\alpha}M\left(\frac{R^2S}{2\alpha},\frac{R^2T}{2\alpha}\right),\label{mffsm}\\
\mathcal{N}:=&\frac{1}{8\pi^{h}}\prod_i\frac{1}{\sqrt{\Gamma(\Delta_i)\Gamma(\Delta_i-h+1)}}, \;\;\; h:=d/2,
\end{align}
where $R$ is the AdS radius, and we use the convention in \cite{Costa:2012cb} and should choose the integration contour to the right of all poles of the integrand. In this formula, $M(s,t)$ are the Mellin amplitudes of four-point functions with single-trace scalar operators in large-$N$ CFTs, and $\mathcal{T}(S,T)$ are the flat space S-matrices of four-point scattering with the dual massless scalar fields. In the AdS/CFT correspondence, the scaling dimensions $\Delta_i$ are related to masses $m_i$ of the bulk scalar fields as
\begin{align}
\Delta_i(\Delta_i-d)=R^2m_i^2,\label{mdf}
\end{align}
and we consider the following massless limit:
\begin{align}
R\to\infty, \;\;\; m_i\to0, \;\;\; R^2m_i^2\sim O(1).
\end{align}
We note that a formula for external massive scalar fields has been proposed in \cite{Paulos:2016fap}.

\subsection{Two scaling behaviors of the Mellin amplitudes}
We would like to relate $A$ in (\ref{sbfsm}) and the scaling behaviors of $M(s,t)$ by using (\ref{mffsm}). One would naively expect that $A'$ in (\ref{rlma}) is the same as $A$ in (\ref{sbfsm}) because we take $S\to\infty$ with fixed $T$. However, before taking the large $S$ limit, we should take the flat space limit $R\to\infty$ and seriously consider the order of the two limits. 

For concrete comparisons, we define $A$ in the scaling behaviors of the Mellin amplitudes as
\begin{align}
M\left(\frac{R^2S}{2\alpha},\frac{R^2T}{2\alpha}\right)\propto S^{A}\;\;\; (S\to\infty, \;  \text{fixed} \; T \; \text{after the flat space limit} \; R\to\infty),\label{fslma}
\end{align}
where ``after the flat space limit" means to keep only the leading terms in $M\left(\frac{R^2S}{2\alpha},\frac{R^2T}{2\alpha}\right)$ at $R\to\infty$, and one can obtain (\ref{sbfsm}) by using (\ref{mffsm}) and (\ref{fslma}).\footnote{Strictly speaking, we also need to consider the $\alpha$-integral. We will comment on it in Subsection \ref{subsec:rfssf}.} Due to the flat space limit $R\to\infty$, $A'$ and $A$ do not have to match, which is similar to the two scaling behaviors of $H(\theta,a)$ and $H(\theta,a)|_{a\to0}$ in Section \ref{review}.

Let us give two examples, which are inspired by the examples in \cite{Chandorkar:2021viw}. The first example is 
\begin{align}
M(s,t)=gR^{-5-2h}\left[s^2+t^4\right],
\end{align}
where $g$ is a coupling constant, and it is easy to see that $A'=2$. When we consider the flat space limit of $M\left(\frac{R^2S}{2\alpha},\frac{R^2T}{2\alpha}\right)$, the second term is dominant, and we conclude that $A=0$ and $A'>A$ in this example. The second one is
\begin{align}
M(s,t)=gR^{3-2h}\left[(s^2+t^4)^{1/2}-t^2\right],\label{sema}
\end{align}
which is an example of $A'<A$ since the scaling exponents are $A'=1$ and $A=2$. 

\section{Inequality \texorpdfstring{$A'\ge A$}{} from polynomial boundedness of \texorpdfstring{$M(s,t)$}{}}\label{pbm}
In this section, for bulk tree-level diagrams with external massless scalar fields corresponding to single-trace operators in CFTs with the conformal block expansion in the $t$-channel, we show that the inequality $A'\ge A$ is derived from criteria, which include polynomial boundedness of the Mellin amplitudes \cite{Fitzpatrick:2012cg}, for CFTs described by perturbative effective field theories in AdS.

\subsection{Review: polynomial boundedness of the Mellin amplitudes}
In the AdS/CFT correspondence, it is a very important problem to determine the conditions for CFTs to have holographic duals. As investigated in \cite{Heemskerk:2009pn, Heemskerk:2010ty, Fitzpatrick:2010zm, El-Showk:2011yvt}, for descriptions by perturbative effective field theories in AdS to be possible, CFTs should admit a perturbative large-$N$ expansion and an approximate  Fock space generated by a finite number of single-trace operators with scaling dimensions below a large gap $\Delta_\Lambda$. In addition to these conditions, the authors of \cite{Fitzpatrick:2012cg} proposed a criterion called polynomial boundedness as follows:
\begin{quote}
The Mellin amplitudes are polynomially bounded at large values of $\delta_{ij}$ below the large gap $\Delta_\Lambda$.
\end{quote}
Simply put, this condition means that conformal correlation functions can be approximately expressed by Witten diagrams in AdS below the scale $\Delta_\Lambda$.

Let us explain this polynomial boundedness at tree-level in detail. As explicitly calculated in \cite{Penedones:2010ue, Fitzpatrick:2011ia}, the Mellin amplitudes for bulk tree-level diagrams with four external scalars are sums over poles and polynomials in the Mandelstam-like variables $s$ and $t$, which satisfy the polynomial boundedness. Conversely, consider conformal correlation functions of four single-trace scalar fields. To the leading order in the large-$N$ expansion, poles of the Mellin amplitudes in $t$ and their residues can be determined from the operator product expansion (OPE) between the finite number of single-trace operators in the $t$-channel \cite{Mack:2009mi, Costa:2012cb, Fitzpatrick:2012cg}. From the polynomial boundedness condition, analytic parts of the Mellin amplitudes except the poles should behave as polynomials at large values of the Mandelstam-like variables, which correspond to the high energy region of the effective field theories in AdS. The corresponding bulk tree-level diagrams can be determined from these poles and polynomials.

\subsection{Inequality \texorpdfstring{$A'\ge A$}{} from the polynomial boundedness}
By using the polynomial boundedness, we derive $A'\ge A$ in the Mellin amplitudes of CFTs at the leading order in $1/N$ that can be described by perturbative effective field theories in AdS. Here, we estimate $A'$ and $A$ in the CFTs with the conformal block expansion in the $t$-channel for a finite number of tree-level contact diagrams and tree-level exchange diagrams below the scale $\Delta_\Lambda$. From the polynomial boundedness and the OPE structure in the $t$-channel, the Mellin amplitudes of four single-trace scalar fields to the leading order in $1/N$ should behave at large values of the Mandelstam-like variables as \cite{Fitzpatrick:2012cg}
\begin{align}
M(s,t)\sim\sum_{\Delta, J, m}\left[\frac{Q_{\Delta, J, m}(s)}{t-\Delta+J-2m}\right]+R(s,t),\label{maalmv}
\end{align}
where $Q_{\Delta, J, m}(s)$ are polynomials in $s$, and $R(s,t)$ is a polynomial in $s$ and $t$. The form (\ref{maalmv}) is similar to tree-level flat space S-matrices given by local Lagrangians, which are the sum of polynomials and a finite number of pole terms for exchange particles \cite{Chowdhury:2019kaq}. The Mellin poles in (\ref{maalmv}) and their residues are determined by OPE coefficients with the exchange of single-trace operators \cite{Mack:2009mi, Costa:2012cb, Fitzpatrick:2012cg}, where $\Delta$ and $J$ are scaling dimensions and spins of the exchange operators, and $m$ is an index for descendants. The degrees of the polynomials in (\ref{maalmv}) are bounded by a finite integer $L$. 

By using a similar analysis as in Subsection \ref{subsec:comp}, we can derive $A'\ge A$ in the Mellin amplitudes (\ref{maalmv}) as follows. The exponent $A'$ is determined by the maximum degree $L$ of the polynomials in (\ref{maalmv}). On the other hand, $A$ does not have to be $L$ due to the flat space limit $R\to\infty$. Since $Q_{\Delta, J, m}(s)$ and $R(s,t)$ are polynomials, $A$ is an integer less than or equal to $L$. Therefore, we obtain $A'=L\ge A$.

We note that the polynomial boundedness constrains the Mellin amplitudes as (\ref{maalmv}) only at large values, not at all values of the Mandelstam-like variables. Similarly, the scaling exponents $A'$ and $A$ are determined from only some regions in the coordinate space of the Mandelstam-like variables.

We comment on a subtlety in the scaling exponents of $M(s,t)$ with the Mellin poles. At $t=\Delta_k-J_k+2m$, $A'$ of (\ref{maalmv}) is not well-defined because $M(s,t)$ diverges. Moreover, if we first take the limit $t\to\Delta_k-J_k+2m$, $A'$ is controlled by $Q_{\Delta_k, J_k, m}(s)$ only. To precisely define the scaling exponents around the poles, one can use a smearing function \cite{Haring:2022cyf, martin1966improvement}. In this paper, instead of such a method, we simply estimate $A'$ and $A$ from the scaling behaviors of $Q_{\Delta, J, m}(s)$ and $R(s,t)$.

To justify our estimation, both $Q_{\Delta, J, m}(s)$ and $R(s,t)$ should contribute to $G^{\text{norm}}(u,v)$ (\ref{nma}) and $\mathcal{T}(S,T)$ (\ref{mffsm}). These polynomials contribute to $G^{\text{norm}}(u,v)$ through the residues of the poles in $M(s,t)$ and of the gamma functions for the $t$-integral in (\ref{nma}). One can also see that both terms in (\ref{maalmv}) contribute to $\mathcal{T}(S,T)$ in the computations of tree-level exchange and contact diagrams \cite{Penedones:2010ue, Fitzpatrick:2011ia}.

To extract the scaling behaviors of tree-level parts in non-perturbative amplitudes, it is convenient to consider the local growth of the amplitudes \cite{Caron-Huot:2017vep, Haring:2022cyf}. In terms of CFTs, the local growth of $G^{\text{norm}}(u,v)$ and $M(s,t)$ in the Regge limits is bounded due to physical assumptions \cite{Caron-Huot:2017vep, Maldacena:2015waa, Penedones:2019tng}. From this bound of $A'$ and the inequality $A'\ge A$, it follows that the scaling exponent $A$ of the flat space S-matrices is bounded. Note that the bound of $A'$ is not necessary for our derivation of the inequality $A'\ge A$.

\section{Inequality \texorpdfstring{$A'\ge A$}{} from the conformal block expansion}\label{icbe}
In the previous section, we did not constrain the details of the polynomials. Assuming the conformal block expansion with intermediate spins bounded above by a finite integer $L$, the polynomials are constrained. In this section, by revisiting the computations in \cite{Costa:2012cb}, we derive an equality $A'=A$ from the conformal block expansion in the $t$-channel with coefficients that are not small in the flat space limit $R\to\infty$. We also explain how $A'>A$ occurs when the coefficients become small in the flat space limit, in terms of CFTs and bulk diagrams.

\subsection{Review: formulas of \texorpdfstring{$M(s,t)$}{} and \texorpdfstring{$\mathcal{T}(S,T)$}{} with the conformal block expansion}\label{subsec:rfssf}
We summarize formulas in \cite{Costa:2012cb} for drawing our conclusion. See Appendix \ref{app} for more details of the convention. Consider the following conformal block expansion of the normalized four-point scalar correlation functions in the $t$-channel with a discrete spectrum:
\begin{align}
G^{\text{norm}}(u,v)=\sum_kC_{12k}C_{34k}G_{\Delta_k, J_k}(u,v),\label{cbe2}
\end{align}
where $G_{\Delta_k, J_k}(u,v)$ are the conformal blocks with the exchange of operators $\mathcal{O}_k$, $\Delta_k$ and $J_k$ are scaling dimensions and spins of $\mathcal{O}_k$, and $C_{ijk}$ are the OPE coefficients. We replace the discrete sum of $\Delta_k$ with an integral representation as
\begin{align}
G^{\text{norm}}(u,v)=\sum_{J=0}^L\int^{\infty}_{-\infty}d\nu b_J(\nu^2)F_{\nu, J}(u,v),\label{cbe}
\end{align}
where $F_{\nu, J}(u,v)$ are the conformal partial waves, which are linear combinations of the conformal blocks and their shadow blocks. Here, we assume that the spins $J_k$ are bounded by a finite integer $L$. The Mellin amplitudes of (\ref{cbe}) are given by
\begin{align}
M(s,t)=\sum_{J=0}^L\int^{\infty}_{-\infty}d\nu b_J(\nu^2)M_{\nu,J}(s,t),\label{cpwema}
\end{align}
where $M_{\nu,J} (s,t)=\omega_{\nu,J}(t)P_{\nu,J}(s,t)$ are the Mellin amplitudes of $F_{\nu, J}(u,v)$. From the normalization of the Mack polynomials $P_{\nu,J}(s,t)\sim s^J+O(s^{J-1})$, the large $s$ behaviors of (\ref{cpwema}) are given by \cite{Costa:2012cb}
\begin{align}
M(s,t)\sim s^L\int^{\infty}_{-\infty}d\nu b_L(\nu^2)\omega_{\nu,L}(t)  \;\;\; (s\to\infty,  \; \text{fixed} \; t). \label{macbe}
\end{align}

The conditions that CFTs have descriptions by holographic effective theories place constraints on the OPE coefficients $C_{ijk}$ and the partial amplitudes $b_J(\nu^2)$. For example, the Mellin amplitude of a single conformal block does not satisfy the polynomial boundedness \cite{Fitzpatrick:2012cg}. To the leading order in $1/N$ of $M(s,t)$ with four external single-trace operators, the partial amplitudes $b_J(\nu^2)$ must have the following poles for single-trace operators $\mathcal{O}_k$ \cite{Costa:2012cb}:
\begin{align}
b_{J_k}(\nu^2)\sim C_{12k}C_{34k}\frac{K_{\Delta_k, J_k}}{\nu^2+(\Delta_k-h)^2}.\label{psbj}
\end{align} 

By substituting (\ref{cpwema}) into (\ref{mffsm}), the authors of \cite{Costa:2012cb} obtained the flat space S-matrices $\mathcal{T}(S,T)$ for external massless scalars in the bulk as
\begin{align}
\mathcal{T}(S,T)=&\sum_{J=0}^LP_{J}(z)a_J(T), \;\;\; z:=1+2S/T,\label{fssmcbe}\\
 a_J(T):=&\frac{1}{\mathcal{N}}\lim_{R\to\infty}R^{2h-3}\left(\frac{R^2T}{4}\right)^J\langle b_J\rangle_T,\label{ajt}\\
 \langle b_J\rangle_T:=&\int dx\delta_{\ell}(x)b_J(-R^2T+x),
\end{align} 
where $P_{J}(z)$ are the partial waves in $(d+1)$-dimensional flat spacetime, and $\delta_{\ell}(x)$ is a regulated delta-function with a length scale $\ell=(-R^2T)^{2/3}$. In their derivation, it is assumed that the following order of limits can be taken. First, take the flat space limit of $M_{\nu,J} (s,t)$ with $|t|\sim|s|\gg|\nu|\gg1$. Next, perform the $\alpha$-integral to obtain the delta-function $\delta_{\ell}(x)$. Finally, perform the $\nu$-integral to obtain the expectation value $\langle b_J\rangle_T$. In particular, the $\alpha$-integral does not change the scaling behaviors of $M_{\nu, J}\left(\frac{R^2S}{2\alpha},\frac{R^2T}{2\alpha}\right)$ with respect to $S$.  If $\langle b_J\rangle_T\sim O(R^{3-2h-2J})$ holds, $a_J(T)$ is well-defined in the flat space limit.

\subsection{Inequality \texorpdfstring{$A'\ge A$}{} from the conformal block expansion in the \texorpdfstring{$t$}{}-channel}
We evaluate the relation between $A'$ and $A$ of (\ref{cpwema}) by comparing (\ref{macbe}) and (\ref{fssmcbe}). From (\ref{macbe}), one can directly obtain $A'=L$. To estimate $A$ from (\ref{fssmcbe}), we have to be careful that $\langle b_J\rangle_T$ depends on $R$. When $P_{L}(z)a_L(T)$ is one of the leading terms in (\ref{fssmcbe}) at $R\to\infty$, we obtain $A'=L=A$. When $P_{L}(z)a_L(T)$ is small compared to the other terms at $R\to\infty$, we obtain $A'=L>A$. Here, we use $P_J(z)\sim z^J+O(z^{J-1})$ and assume that $a_J(T)$ is well-defined in the flat space limit.

We discuss a mechanism by which $a_J(T)$ becomes small from the viewpoint of the $R$-dependence of $\langle b_J\rangle_T$.
To see the $R$-dependence of $\langle b_J\rangle_T$, let us focus on the pole structure (\ref{psbj}) of $b_{J}(\nu^2)$. From the definition of $K_{\Delta,J}$ (\ref{kdj}), one can obtain
\begin{align}
K_{\Delta, J+2}\sim\frac{16}{\Delta^4}K_{\Delta,J}\propto \frac{1}{R^4}K_{\Delta,J} \;\;\; (R\to\infty),
\end{align}
where we use $\Delta_i\sim O(1)$ for external massless scalars and $\Delta\propto R$. As the spin $J$ becomes larger, $K_{\Delta,J}$ becomes smaller at $R\to\infty$. This decrease is offset by the increase of $R^{2J}$ in (\ref{ajt}) like $R^{2(J+2)}K_{\Delta, J+2}\propto R^{2J}K_{\Delta,J}$. The OPE coefficients $C_{ijk}$ in (\ref{psbj}) may depend on the scaling dimensions of $\mathcal{O}_k$ and hence $R$. When $C_{ijk}$ for the operators $\mathcal{O}_k$ with the spin $J$ are small at $R\to\infty$, the smallness of $C_{ijk}$ can lead to the smallness of $\langle b_J\rangle_T$ and $a_J(T)$ at $R\to\infty$.

We also give an intuitive explanation of $A'>A$ from the smallness of $C_{ijk}$ in (\ref{cbe2}) at $R\to\infty$. In the Regge limit of $G^{\text{norm}}(u,v)$ on the Causally Regge sheet, the conformal blocks with the maximum spin $J_k=L$ are dominant in the conformal block expansion (\ref{cbe2}) \cite{Costa:2012cb, Cornalba:2006xk, Cornalba:2006xm, Cornalba:2007zb, Cornalba:2007fs, Cornalba:2008qf}. When $C_{ijk}$ for the operators with the maximum spin $L$ are small at $R\to\infty$, the contributions from $G_{\Delta_k, L}(u,v)$ are ignored in the flat space limit, and the scaling exponent $A$ in the flat space limit is smaller than $A'$ like $A'=L>A$.

Another example by tree-level Witten diagrams is as follows. Consider a four-point contact Witten diagram with an interaction vertex $g_0\phi_1\phi_2\phi_3\phi_4$, where $g_0$ is a coupling constant, and $\phi_i$ are bulk scalar fields dual to $\mathcal{O}_i$ in CFTs. The Mellin amplitude $M_0(s,t)$ of this Witten diagram is given by \cite{Penedones:2010ue}
\begin{align}
M_0(s,t)=g_0\,\mathcal{N}\,\Gamma\left(\frac{1}{2}\sum_i\Delta_i-h\right)R^{3-2h},
\end{align}
where our normalization of the Mellin amplitude is different from the normalization in \cite{Penedones:2010ue}. We also consider another contact Witten diagram with an interaction vertex that includes a coupling constant $g_L$ and $L$ pairs of contracted derivatives acting on $\phi_1$ and $\phi_3$. For example, the vertex with $L=1$ is given by $g_1(\nabla_A\phi_1)\phi_2(\nabla^A\phi_3)\phi_4$, where $\nabla_A$ is the covariant derivative. In the large $\delta_{ij}$ limit, the Mellin amplitude $M_L(s,t)$ of this Witten diagram with the derivative interaction is given by \cite{Penedones:2010ue}
\begin{align}
M_L(s,t)\sim g_L\,\mathcal{N}\,\Gamma\left(\frac{1}{2}\sum_i\Delta_i-h+L\right)R^{3-2h-2L}s^L.\label{mawddi}
\end{align}
One can also obtain the Mellin amplitude depending on $t$ by considering other contractions of the covariant derivatives.
For the conformal block expansion of contact Witten diagrams, see \cite{Heemskerk:2009pn, Hijano:2015zsa, Bekaert:2015tva}.

We analyze the scaling exponents $A'$ and $A$ of $M(s,t)=M_0(s,t)+M_L(s,t)$. From (\ref{mawddi}), the scaling exponent $A'$ is given by $A'=L$. Suppose that $g_0$ and $g_L$ do not depend on $R$. Then, $M_0\left(\frac{R^2S}{2\alpha},\frac{R^2T}{2\alpha}\right)$ and $M_L\left(\frac{R^2S}{2\alpha},\frac{R^2T}{2\alpha}\right)$ are proportional to $R^{3-2h}$ in the flat space limit, and this $R$-dependence is canceled by $R^{2h-3}$ in (\ref{mffsm}). In this case, we obtain $A'=L=A$. On the other hand, when $g_L$ is small at $R\to\infty$ such as $g_L\sim O(R^{-1})$, $M_L\left(\frac{R^2S}{2\alpha},\frac{R^2T}{2\alpha}\right)$ does not contribute to $\mathcal{T}(S,T)$, and therefore we obtain $A'=L>0=A$.

\section{Summary and discussion}\label{summary}
We investigated relation between two scaling exponents $A'$ and $A$ in Mellin amplitudes of four-point functions with single-trace scalars. These exponents are related to scaling behaviors in the Regge limits of normalized four-point functions $G^{\text{norm}}(u,v)$ and flat space S-matrices $\mathcal{T}(S,T)$, respectively. The four-point functions we studied admit the conformal block expansion in the $t$-channel with the maximum intermediate spin $L$ for tree-level bulk diagrams with four external massless scalars. We showed that an inequality $A'\ge A$ is derived from polynomial boundedness of the Mellin amplitudes, which is a criterion for descriptions of CFTs by perturbative effective field theories in the bulk. We also analyzed the inequality by using a formula of $\mathcal{T}(S,T)$ with the conformal block expansion and derived an equality $A'=A$ when coefficients are not small in the flat space limit.
Assuming the bound of $A'$, the inequality $A'\ge A$ leads to the bound of $A$ for a finite number of tree-level contact and exchange diagrams in the $t$-channel.

Finally, we discuss some future directions of our work.

\paragraph{Inequality from expansions in the other channels}
Assuming crossing symmetry, we can perform the conformal block expansion of $G^{\text{norm}}(u,v)$ in the $s$-channel and the $u$-channel. The Polyakov blocks, which correspond to tree-level Witten exchange diagrams \cite{Polyakov:1974gs, Gopakumar:2016wkt, Gopakumar:2016cpb},  can be also used to expand $G^{\text{norm}}(u,v)$, and the Mellin amplitudes of the Polyakov blocks have poles in the Mandelstam-like variables (see, for instance, \cite{Gopakumar:2018xqi, Mazac:2019shk, Sleight:2019ive, Gopakumar:2021dvg}). It is interesting to study the inequality from these expansions for the tree-level exchange diagrams.

\paragraph{Inequality at \texorpdfstring{$L\to\infty$}{}}
In the string theory, there are infinitely many higher spin fields, and we need to take $L\to\infty$. The conformal Regge theory for the resummation of higher spin fields at $L\to\infty$  was developed by \cite{Costa:2012cb}. It is important to study the inequality at $L\to\infty$, paying attention to the order of the large $S$ limit and the flat space limit $R\to\infty$.

\paragraph{Inequality with external spinning fields}
Toward application to the CRG conjecture with external photons and gravitons, it is significant to generalize our result to the Mellin amplitudes of external spinning fields  \cite{Paulos:2011ie, Goncalves:2014rfa, Faller:2017hyt, Chen:2017xdz, Sleight:2018epi, Silva:2021ece}.  Recent developments in the flat space limit of the AdS/CFT correspondence for external spinning fields \cite{Raju:2012zr, Hijano:2020szl, Antunes:2020pof, Caron-Huot:2021kjy, Li:2021snj, Jain:2022ujj} will be helpful in the generalization. 

\paragraph{Inequality at loop-level}
Recently, subleading chaos bounds have been proposed \cite{Kundu:2021qcx, Kundu:2021mex}, where these bounds constrain the local growth of subleading terms in the large $N$ expansion of $G^{\text{norm}}(u,v)$ in the Regge limit. It is interesting to investigate the inequality for loop-level diagrams because the subleading chaos bounds lead to the local growth bounds of $\mathcal{T}(S,T)$ at loop-level from the inequality. The Mellin amplitudes are also useful for the analysis of bulk loop-level diagrams \cite{Penedones:2010ue, Fitzpatrick:2011hu, Fitzpatrick:2012cg}.

\acknowledgments

We would like to thank members of the laboratory in GIST for valuable discussions and comments. In particular, we would like to thank Keun-Young Kim for reading and commenting on the manuscript.
This work was supported by Basic Science Research Program through the National Research Foundation of Korea (NRF) funded by the Ministry of Science, ICT \& Future
Planning (NRF-2021R1A2C1006791) and the Ministry of Education (NRF-2020R1I1A1A01072726), and by the GIST Research Institute(GRI) grant funded by GIST in 2022.

\appendix

\section{Definitions of the functions}\label{app}
In this appendix, we summarize definitions of the functions used in \cite{Costa:2012cb}.

The conformal blocks $G_{\Delta, J}(u,v)$ are solutions of the conformal Casimir equation \cite{Dolan:2003hv}
\begin{align}
\begin{aligned}
\big[&(1-u-v)\partial_v(v\partial_v-\Delta_{12}/2+\Delta_{34}/2)+u\partial_u(2u\partial_u-d)\\
&-(1+u-v)(u\partial_u+v\partial_v-\Delta_{12}/2)(u\partial_u+v\partial_v+\Delta_{34}/2)\big]G_{\Delta, J}(u,v)\\
=&\frac{1}{2}[\Delta(\Delta-d)+J(J+d-2)]G_{\Delta, J}(u,v),
\end{aligned}
\end{align}
with the following boundary condition:
\begin{align}
G_{\Delta, J}(u,v)\sim\frac{J!}{2^J(h-1)_J}u^{\Delta/2}C_{J}^{h-1}\left(\frac{v-1}{2\sqrt{u}}\right) \;\;\; \left(u\to0, v\to1, \text{fixed} \; \frac{v-1}{\sqrt{u}}\right),
\end{align}
where $(x)_n:=\Gamma(x+n)/\Gamma(x)$ is the Pochhammer symbol, and $C_n^{\alpha}(x)$ is the Gegenbauer polynomial
\begin{align}
C_n^{\alpha}(x):=\frac{(-1)^n}{2^n n!}\frac{\Gamma(\alpha+1/2)\Gamma(n+2\alpha)}{\Gamma(2\alpha)\Gamma(\alpha+n+1/2)}(1-x^2)^{-\alpha+1/2}\frac{d^n}{d x^n}[(1-x^2)^{n+\alpha-1/2}].
\end{align}
The conformal partial waves $F_{\nu, J}(u,v)$ are defined as
\begin{align}
F_{\nu, J}(u,v):=\kappa_{\nu,J}G_{h+i\nu,J}(u,v)+\kappa_{-\nu,J}G_{h-i\nu,J}(u,v), \;\;\; \kappa_{\nu,J}:=\frac{i\nu}{2\pi K_{h+i\nu,J}},\\
\begin{aligned}
K_{\Delta,J}:=&\frac{\Gamma(\Delta+J)\Gamma(\Delta-h+1)(\Delta-1)_J}{4^{J-1}\Gamma\left(\frac{\Delta+J+\Delta_{12}}{2}\right)\Gamma\left(\frac{\Delta+J-\Delta_{12}}{2}\right)\Gamma\left(\frac{\Delta+J+\Delta_{34}}{2}\right)\Gamma\left(\frac{\Delta+J-\Delta_{34}}{2}\right)}\\
\times&\frac{1}{\Gamma\left(\frac{\Delta_1+\Delta_2-\Delta+J}{2}\right)\Gamma\left(\frac{\Delta_3+\Delta_4-\Delta+J}{2}\right)\Gamma\left(\frac{\Delta_1+\Delta_2+\Delta+J-d}{2}\right)\Gamma\left(\frac{\Delta_3+\Delta_4+\Delta+J-d}{2}\right)}.\label{kdj}
\end{aligned}
\end{align}

The Mellin amplitude $M_{\nu,J}(s,t)$ of $F_{\nu, J}(u,v)$ is given by
\begin{align}
M_{\nu,J}(s,t)=&\,\omega_{\nu,J}(t)P_{\nu,J}(s,t),\\
\omega_{\nu,J}(t):=&\frac{\Gamma\left(\frac{\Delta_1+\Delta_2+J+i\nu-h}{2}\right)\Gamma\left(\frac{\Delta_3+\Delta_4+J+i\nu-h}{2}\right)\Gamma\left(\frac{\Delta_1+\Delta_2+J-i\nu-h}{2}\right)\Gamma\left(\frac{\Delta_3+\Delta_4+J-i\nu-h}{2}\right)}{8\pi\Gamma(i\nu)\Gamma(-i\nu)}\notag\\
\times&\frac{\Gamma\left(\frac{h+i\nu-J-t}{2}\right)\Gamma\left(\frac{h-i\nu-J-t}{2}\right)}{\Gamma\left(\frac{\Delta_1+\Delta_2-t}{2}\right)\Gamma\left(\frac{\Delta_3+\Delta_4-t}{2}\right)},\\
P_{\nu,J}(s,t):=&\sum_{r=0}^{[J/2]}a_{J,r}\frac{2^{J+2r}\left(\frac{h+i\nu-J-t}{2}\right)_r\left(\frac{h-i\nu-J-t}{2}\right)_r(J-2r)!}{(h+i\nu-1)_J(h-i\nu-1)_J}\notag\\
\times&\sum_{\{k_{ij}\}:\sum' k_{ij}=J-2r}(-1)^{k_{13}+k_{24}}\prod'\frac{(\delta_{ij})_{k_{ij}}}{k_{ij}!}\prod_{n=1}^4(\alpha_n)_{J-r-\sum_jk_{jn}},
\end{align}
where $[J/2]$ is the floor function of $J/2$.
In the Mack polynomial $P_{\nu,J}(s,t)$, $\sum'$ and $\prod'$ are over four labels $(ij)=(13), (14), (23),(24)$, and $k_{ij} \, (i\ne j)$ are nonnegative integers with $k_{ij}=k_{ji}$. The Mellin variables $\delta_{ij}$ are related to $s$ and $t$ as
\begin{align}
\delta_{13}=\frac{\Delta_{34}-s}{2}, \;\;\; \delta_{14}=\frac{t+s+\Delta_{12}-\Delta_{34}}{2}, \;\;\; \delta_{23}=\frac{t+s}{2}, \;\;\; \delta_{24}=-\frac{\Delta_{12}+s}{2}.
\end{align}
The variables $\alpha_n$ are defined by
\begin{align}
\begin{aligned}
\alpha_1:=1-\frac{h+i\nu+J+\Delta_{12}}{2}, \;\;\; \alpha_2:=1-\frac{h+i\nu+J-\Delta_{12}}{2},\\
\alpha_3:=1-\frac{h-i\nu+J+\Delta_{34}}{2}, \;\;\; \alpha_2:=1-\frac{h-i\nu+J-\Delta_{12}}{2},
\end{aligned}
\end{align}
and the coefficients $a_{J,r}$ are defined by
\begin{align}
a_{J,r}:=(-1)^r\frac{J!(h+J-1)_{-r}}{2^{2r}r!(J-2r)!}.
\end{align}
By using $a_{J,r}$, we define the partial waves $P_J(z)$ in $(d+1)$-dimensional flat spacetime as
\begin{align}
P_J(z):=\sum_{r=0}^{[J/2]}a_{J,r}z^{J-2r}.
\end{align}

A regularized delta-function $\delta_\ell(x)$ with $\ell=(-R^2T)^{2/3}$ is defined by
\begin{align}
\delta_\ell\left(1+\frac{\nu^2}{R^2T}\right):=\int^{i\infty}_{-i\infty}\frac{d \alpha}{2\pi i}\exp\left[\alpha+\nu\arctan\left(\frac{2\alpha\nu}{R^2T}\right)-\frac{R^2T}{4\alpha}\log\left(1+\left(\frac{2\alpha\nu}{R^2T}\right)^2\right)\right]
\end{align}
If we consider only the first order term with respect to $\alpha$ in the exponential, $\delta_\ell(x)$ is approximated by the usual delta-function $\delta(x):=\int^{i\infty}_{-i\infty}\frac{d \alpha}{2\pi i}e^{\alpha x}$.

\bibliography{HyunSikRefs}

\providecommand{\href}[2]{#2}\begingroup\raggedright\begin{thebibliography}{10}

\bibitem{Chowdhury:2019kaq}
S.~D. Chowdhury, A.~Gadde, T.~Gopalka, I.~Halder, L.~Janagal and S.~Minwalla,
  \emph{{Classifying and constraining local four photon and four graviton
  S-matrices}}, \href{http://dx.doi.org/10.1007/JHEP02(2020)114}{\emph{JHEP}
  {\bf 02} (2020) 114}, [\href{http://arxiv.org/abs/1910.14392}{{\tt
  1910.14392}}].

\bibitem{Chakraborty:2020rxf}
S.~Chakraborty, S.~D. Chowdhury, T.~Gopalka, S.~Kundu, S.~Minwalla and
  A.~Mishra, \emph{{Classification of all 3 particle S-matrices quadratic in
  photons or gravitons}},
  \href{http://dx.doi.org/10.1007/JHEP04(2020)110}{\emph{JHEP} {\bf 04} (2020)
  110}, [\href{http://arxiv.org/abs/2001.07117}{{\tt 2001.07117}}].

\bibitem{Haring:2022cyf}
K.~H\"aring and A.~Zhiboedov, \emph{{Gravitational Regge bounds}},
  \href{http://arxiv.org/abs/2202.08280}{{\tt 2202.08280}}.

\bibitem{Maldacena:2015waa}
J.~Maldacena, S.~H. Shenker and D.~Stanford, \emph{{A bound on chaos}},
  \href{http://dx.doi.org/10.1007/JHEP08(2016)106}{\emph{JHEP} {\bf 08} (2016)
  106}, [\href{http://arxiv.org/abs/1503.01409}{{\tt 1503.01409}}].

\bibitem{Cornalba:2006xk}
L.~Cornalba, M.~S. Costa, J.~Penedones and R.~Schiappa, \emph{{Eikonal
  Approximation in AdS/CFT: From Shock Waves to Four-Point Functions}},
  \href{http://dx.doi.org/10.1088/1126-6708/2007/08/019}{\emph{JHEP} {\bf 08}
  (2007) 019}, [\href{http://arxiv.org/abs/hep-th/0611122}{{\tt
  hep-th/0611122}}].

\bibitem{Cornalba:2006xm}
L.~Cornalba, M.~S. Costa, J.~Penedones and R.~Schiappa, \emph{{Eikonal
  Approximation in AdS/CFT: Conformal Partial Waves and Finite N Four-Point
  Functions}},
  \href{http://dx.doi.org/10.1016/j.nuclphysb.2007.01.007}{\emph{Nucl. Phys.}
  {\bf B767} (2007) 327--351}, [\href{http://arxiv.org/abs/hep-th/0611123}{{\tt
  hep-th/0611123}}].

\bibitem{Cornalba:2007zb}
L.~Cornalba, M.~S. Costa and J.~Penedones, \emph{{Eikonal approximation in
  AdS/CFT: Resumming the gravitational loop expansion}},
  \href{http://dx.doi.org/10.1088/1126-6708/2007/09/037}{\emph{JHEP} {\bf 09}
  (2007) 037}, [\href{http://arxiv.org/abs/0707.0120}{{\tt 0707.0120}}].

\bibitem{Cornalba:2007fs}
L.~Cornalba, \emph{{Eikonal methods in AdS/CFT: Regge theory and multi-reggeon
  exchange}},  \href{http://arxiv.org/abs/0710.5480}{{\tt 0710.5480}}.

\bibitem{Cornalba:2008qf}
L.~Cornalba, M.~S. Costa and J.~Penedones, \emph{{Eikonal Methods in AdS/CFT:
  BFKL Pomeron at Weak Coupling}},
  \href{http://dx.doi.org/10.1088/1126-6708/2008/06/048}{\emph{JHEP} {\bf 06}
  (2008) 048}, [\href{http://arxiv.org/abs/0801.3002}{{\tt 0801.3002}}].

\bibitem{Shenker:2013pqa}
S.~H. Shenker and D.~Stanford, \emph{{Black holes and the butterfly effect}},
  \href{http://dx.doi.org/10.1007/JHEP03(2014)067}{\emph{JHEP} {\bf 03} (2014)
  067}, [\href{http://arxiv.org/abs/1306.0622}{{\tt 1306.0622}}].

\bibitem{Roberts:2014ifa}
D.~A. Roberts and D.~Stanford, \emph{{Two-dimensional conformal field theory
  and the butterfly effect}},
  \href{http://dx.doi.org/10.1103/PhysRevLett.115.131603}{\emph{Phys. Rev.
  Lett.} {\bf 115} (2015) 131603}, [\href{http://arxiv.org/abs/1412.5123}{{\tt
  1412.5123}}].

\bibitem{Shenker:2014cwa}
S.~H. Shenker and D.~Stanford, \emph{{Stringy effects in scrambling}},
  \href{http://dx.doi.org/10.1007/JHEP05(2015)132}{\emph{JHEP} {\bf 05} (2015)
  132}, [\href{http://arxiv.org/abs/1412.6087}{{\tt 1412.6087}}].

\bibitem{Perlmutter:2016pkf}
E.~Perlmutter, \emph{{Bounding the Space of Holographic CFTs with Chaos}},
  \href{http://dx.doi.org/10.1007/JHEP10(2016)069}{\emph{JHEP} {\bf 10} (2016)
  069}, [\href{http://arxiv.org/abs/1602.08272}{{\tt 1602.08272}}].

\bibitem{Ahn:2019rnq}
Y.~Ahn, V.~Jahnke, H.-S. Jeong and K.-Y. Kim, \emph{{Scrambling in Hyperbolic
  Black Holes: shock waves and pole-skipping}},
  \href{http://dx.doi.org/10.1007/JHEP10(2019)257}{\emph{JHEP} {\bf 10} (2019)
  257}, [\href{http://arxiv.org/abs/1907.08030}{{\tt 1907.08030}}].

\bibitem{Gary:2009ae}
M.~Gary, S.~B. Giddings and J.~Penedones, \emph{{Local bulk S-matrix elements
  and CFT singularities}},
  \href{http://dx.doi.org/10.1103/PhysRevD.80.085005}{\emph{Phys. Rev. D} {\bf
  80} (2009) 085005}, [\href{http://arxiv.org/abs/0903.4437}{{\tt 0903.4437}}].

\bibitem{Heemskerk:2009pn}
I.~Heemskerk, J.~Penedones, J.~Polchinski and J.~Sully, \emph{{Holography from
  Conformal Field Theory}},
  \href{http://dx.doi.org/10.1088/1126-6708/2009/10/079}{\emph{JHEP} {\bf 10}
  (2009) 079}, [\href{http://arxiv.org/abs/0907.0151}{{\tt 0907.0151}}].

\bibitem{Okuda:2010ym}
T.~Okuda and J.~Penedones, \emph{{String scattering in flat space and a scaling
  limit of Yang-Mills correlators}},
  \href{http://dx.doi.org/10.1103/PhysRevD.83.086001}{\emph{Phys. Rev. D} {\bf
  83} (2011) 086001}, [\href{http://arxiv.org/abs/1002.2641}{{\tt 1002.2641}}].

\bibitem{Penedones:2010ue}
J.~Penedones, \emph{{Writing CFT correlation functions as AdS scattering
  amplitudes}}, \href{http://dx.doi.org/10.1007/JHEP03(2011)025}{\emph{JHEP}
  {\bf 03} (2011) 025}, [\href{http://arxiv.org/abs/1011.1485}{{\tt
  1011.1485}}].

\bibitem{Maldacena:2015iua}
J.~Maldacena, D.~Simmons-Duffin and A.~Zhiboedov, \emph{{Looking for a bulk
  point}}, \href{http://dx.doi.org/10.1007/JHEP01(2017)013}{\emph{JHEP} {\bf
  01} (2017) 013}, [\href{http://arxiv.org/abs/1509.03612}{{\tt 1509.03612}}].

\bibitem{Chandorkar:2021viw}
D.~Chandorkar, S.~D. Chowdhury, S.~Kundu and S.~Minwalla, \emph{{Bounds on
  Regge growth of flat space scattering from bounds on chaos}},
  \href{http://dx.doi.org/10.1007/JHEP05(2021)143}{\emph{JHEP} {\bf 05} (2021)
  143}, [\href{http://arxiv.org/abs/2102.03122}{{\tt 2102.03122}}].

\bibitem{Mack:2009mi}
G.~Mack, \emph{{D-independent representation of Conformal Field Theories in D
  dimensions via transformation to auxiliary Dual Resonance Models. Scalar
  amplitudes}},  \href{http://arxiv.org/abs/0907.2407}{{\tt 0907.2407}}.

\bibitem{Costa:2012cb}
M.~S. Costa, V.~Goncalves and J.~Penedones, \emph{{Conformal Regge theory}},
  \href{http://dx.doi.org/10.1007/JHEP12(2012)091}{\emph{JHEP} {\bf 12} (2012)
  091}, [\href{http://arxiv.org/abs/1209.4355}{{\tt 1209.4355}}].

\bibitem{Fitzpatrick:2012cg}
A.~L. Fitzpatrick and J.~Kaplan, \emph{{AdS Field Theory from Conformal Field
  Theory}}, \href{http://dx.doi.org/10.1007/JHEP02(2013)054}{\emph{JHEP} {\bf
  02} (2013) 054}, [\href{http://arxiv.org/abs/1208.0337}{{\tt 1208.0337}}].

\bibitem{Caron-Huot:2021enk}
S.~Caron-Huot, D.~Mazac, L.~Rastelli and D.~Simmons-Duffin, \emph{{AdS bulk
  locality from sharp CFT bounds}},
  \href{http://dx.doi.org/10.1007/JHEP11(2021)164}{\emph{JHEP} {\bf 11} (2021)
  164}, [\href{http://arxiv.org/abs/2106.10274}{{\tt 2106.10274}}].

\bibitem{Witten:1998qj}
E.~Witten, \emph{{Anti-de Sitter space and holography}}, {\emph{Adv. Theor.
  Math. Phys.} {\bf 2} (1998) 253--291},
  [\href{http://arxiv.org/abs/hep-th/9802150}{{\tt hep-th/9802150}}].

\bibitem{Fitzpatrick:2011hu}
A.~L. Fitzpatrick and J.~Kaplan, \emph{{Analyticity and the Holographic
  S-Matrix}}, \href{http://dx.doi.org/10.1007/JHEP10(2012)127}{\emph{JHEP} {\bf
  10} (2012) 127}, [\href{http://arxiv.org/abs/1111.6972}{{\tt 1111.6972}}].

\bibitem{Paulos:2016fap}
M.~F. Paulos, J.~Penedones, J.~Toledo, B.~C. van Rees and P.~Vieira, \emph{{The
  S-matrix bootstrap. Part I: QFT in AdS}},
  \href{http://dx.doi.org/10.1007/JHEP11(2017)133}{\emph{JHEP} {\bf 11} (2017)
  133}, [\href{http://arxiv.org/abs/1607.06109}{{\tt 1607.06109}}].

\bibitem{Heemskerk:2010ty}
I.~Heemskerk and J.~Sully, \emph{{More Holography from Conformal Field
  Theory}}, \href{http://dx.doi.org/10.1007/JHEP09(2010)099}{\emph{JHEP} {\bf
  09} (2010) 099}, [\href{http://arxiv.org/abs/1006.0976}{{\tt 1006.0976}}].

\bibitem{Fitzpatrick:2010zm}
A.~L. Fitzpatrick, E.~Katz, D.~Poland and D.~Simmons-Duffin, \emph{{Effective
  Conformal Theory and the Flat-Space Limit of AdS}},
  \href{http://dx.doi.org/10.1007/JHEP07(2011)023}{\emph{JHEP} {\bf 07} (2011)
  023}, [\href{http://arxiv.org/abs/1007.2412}{{\tt 1007.2412}}].

\bibitem{El-Showk:2011yvt}
S.~El-Showk and K.~Papadodimas, \emph{{Emergent Spacetime and Holographic
  CFTs}}, \href{http://dx.doi.org/10.1007/JHEP10(2012)106}{\emph{JHEP} {\bf 10}
  (2012) 106}, [\href{http://arxiv.org/abs/1101.4163}{{\tt 1101.4163}}].

\bibitem{Fitzpatrick:2011ia}
A.~L. Fitzpatrick, J.~Kaplan, J.~Penedones, S.~Raju and B.~C. van Rees,
  \emph{{A Natural Language for AdS/CFT Correlators}},
  \href{http://dx.doi.org/10.1007/JHEP11(2011)095}{\emph{JHEP} {\bf 11} (2011)
  095}, [\href{http://arxiv.org/abs/1107.1499}{{\tt 1107.1499}}].

\bibitem{martin1966improvement}
A.~Martin, \emph{Improvement of the greenberg-low bound}, {\emph{Il Nuovo
  Cimento A (1965-1970)} {\bf 42} (1966) 901--907}.

\bibitem{Caron-Huot:2017vep}
S.~Caron-Huot, \emph{{Analyticity in Spin in Conformal Theories}},
  \href{http://dx.doi.org/10.1007/JHEP09(2017)078}{\emph{JHEP} {\bf 09} (2017)
  078}, [\href{http://arxiv.org/abs/1703.00278}{{\tt 1703.00278}}].

\bibitem{Penedones:2019tng}
J.~Penedones, J.~A. Silva and A.~Zhiboedov, \emph{{Nonperturbative Mellin
  Amplitudes: Existence, Properties, Applications}},
  \href{http://dx.doi.org/10.1007/JHEP08(2020)031}{\emph{JHEP} {\bf 08} (2020)
  031}, [\href{http://arxiv.org/abs/1912.11100}{{\tt 1912.11100}}].

\bibitem{Hijano:2015zsa}
E.~Hijano, P.~Kraus, E.~Perlmutter and R.~Snively, \emph{{Witten Diagrams
  Revisited: The AdS Geometry of Conformal Blocks}},
  \href{http://dx.doi.org/10.1007/JHEP01(2016)146}{\emph{JHEP} {\bf 01} (2016)
  146}, [\href{http://arxiv.org/abs/1508.00501}{{\tt 1508.00501}}].

\bibitem{Bekaert:2015tva}
X.~Bekaert, J.~Erdmenger, D.~Ponomarev and C.~Sleight, \emph{{Quartic AdS
  Interactions in Higher-Spin Gravity from Conformal Field Theory}},
  \href{http://dx.doi.org/10.1007/JHEP11(2015)149}{\emph{JHEP} {\bf 11} (2015)
  149}, [\href{http://arxiv.org/abs/1508.04292}{{\tt 1508.04292}}].

\bibitem{Polyakov:1974gs}
A.~M. Polyakov, \emph{{Nonhamiltonian approach to conformal quantum field
  theory}}, {\emph{Zh. Eksp. Teor. Fiz.} {\bf 66} (1974) 23--42}.

\bibitem{Gopakumar:2016wkt}
R.~Gopakumar, A.~Kaviraj, K.~Sen and A.~Sinha, \emph{{Conformal Bootstrap in
  Mellin Space}},
  \href{http://dx.doi.org/10.1103/PhysRevLett.118.081601}{\emph{Phys. Rev.
  Lett.} {\bf 118} (2017) 081601}, [\href{http://arxiv.org/abs/1609.00572}{{\tt
  1609.00572}}].

\bibitem{Gopakumar:2016cpb}
R.~Gopakumar, A.~Kaviraj, K.~Sen and A.~Sinha, \emph{{A Mellin space approach
  to the conformal bootstrap}},
  \href{http://dx.doi.org/10.1007/JHEP05(2017)027}{\emph{JHEP} {\bf 05} (2017)
  027}, [\href{http://arxiv.org/abs/1611.08407}{{\tt 1611.08407}}].

\bibitem{Gopakumar:2018xqi}
R.~Gopakumar and A.~Sinha, \emph{{On the Polyakov-Mellin bootstrap}},
  \href{http://dx.doi.org/10.1007/JHEP12(2018)040}{\emph{JHEP} {\bf 12} (2018)
  040}, [\href{http://arxiv.org/abs/1809.10975}{{\tt 1809.10975}}].

\bibitem{Mazac:2019shk}
D.~Maz\'a\v{c}, L.~Rastelli and X.~Zhou, \emph{{A basis of analytic functionals
  for CFTs in general dimension}},
  \href{http://dx.doi.org/10.1007/JHEP08(2021)140}{\emph{JHEP} {\bf 08} (2021)
  140}, [\href{http://arxiv.org/abs/1910.12855}{{\tt 1910.12855}}].

\bibitem{Sleight:2019ive}
C.~Sleight and M.~Taronna, \emph{{The Unique Polyakov Blocks}},
  \href{http://dx.doi.org/10.1007/JHEP11(2020)075}{\emph{JHEP} {\bf 11} (2020)
  075}, [\href{http://arxiv.org/abs/1912.07998}{{\tt 1912.07998}}].

\bibitem{Gopakumar:2021dvg}
R.~Gopakumar, A.~Sinha and A.~Zahed, \emph{{Crossing Symmetric Dispersion
  Relations for Mellin Amplitudes}},
  \href{http://dx.doi.org/10.1103/PhysRevLett.126.211602}{\emph{Phys. Rev.
  Lett.} {\bf 126} (2021) 211602}, [\href{http://arxiv.org/abs/2101.09017}{{\tt
  2101.09017}}].

\bibitem{Paulos:2011ie}
M.~F. Paulos, \emph{{Towards Feynman rules for Mellin amplitudes}},
  \href{http://dx.doi.org/10.1007/JHEP10(2011)074}{\emph{JHEP} {\bf 10} (2011)
  074}, [\href{http://arxiv.org/abs/1107.1504}{{\tt 1107.1504}}].

\bibitem{Goncalves:2014rfa}
V.~Gon\c{c}alves, J.~a. Penedones and E.~Trevisani, \emph{{Factorization of
  Mellin amplitudes}},
  \href{http://dx.doi.org/10.1007/JHEP10(2015)040}{\emph{JHEP} {\bf 10} (2015)
  040}, [\href{http://arxiv.org/abs/1410.4185}{{\tt 1410.4185}}].

\bibitem{Faller:2017hyt}
J.~Faller, S.~Sarkar and M.~Verma, \emph{{Mellin Amplitudes for Fermionic
  Conformal Correlators}},
  \href{http://dx.doi.org/10.1007/JHEP03(2018)106}{\emph{JHEP} {\bf 03} (2018)
  106}, [\href{http://arxiv.org/abs/1711.07929}{{\tt 1711.07929}}].

\bibitem{Chen:2017xdz}
H.-Y. Chen, E.-J. Kuo and H.~Kyono, \emph{{Towards Spinning Mellin
  Amplitudes}},
  \href{http://dx.doi.org/10.1016/j.nuclphysb.2018.04.019}{\emph{Nucl. Phys. B}
  {\bf 931} (2018) 291--323}, [\href{http://arxiv.org/abs/1712.07991}{{\tt
  1712.07991}}].

\bibitem{Sleight:2018epi}
C.~Sleight and M.~Taronna, \emph{{Spinning Mellin Bootstrap: Conformal Partial
  Waves, Crossing Kernels and Applications}},
  \href{http://dx.doi.org/10.1002/prop.201800038}{\emph{Fortsch. Phys.} {\bf
  66} (2018) 1800038}, [\href{http://arxiv.org/abs/1804.09334}{{\tt
  1804.09334}}].

\bibitem{Silva:2021ece}
J.~A. Silva, \emph{{Four point functions in CFT\textquoteright{}s with slightly
  broken higher spin symmetry}},
  \href{http://dx.doi.org/10.1007/JHEP05(2021)097}{\emph{JHEP} {\bf 05} (2021)
  097}, [\href{http://arxiv.org/abs/2103.00275}{{\tt 2103.00275}}].

\bibitem{Raju:2012zr}
S.~Raju, \emph{{New Recursion Relations and a Flat Space Limit for AdS/CFT
  Correlators}},
  \href{http://dx.doi.org/10.1103/PhysRevD.85.126009}{\emph{Phys. Rev. D} {\bf
  85} (2012) 126009}, [\href{http://arxiv.org/abs/1201.6449}{{\tt 1201.6449}}].

\bibitem{Hijano:2020szl}
E.~Hijano and D.~Neuenfeld, \emph{{Soft photon theorems from CFT Ward identites
  in the flat limit of AdS/CFT}},
  \href{http://dx.doi.org/10.1007/JHEP11(2020)009}{\emph{JHEP} {\bf 11} (2020)
  009}, [\href{http://arxiv.org/abs/2005.03667}{{\tt 2005.03667}}].

\bibitem{Antunes:2020pof}
A.~Antunes, M.~S. Costa, T.~Hansen, A.~Salgarkar and S.~Sarkar, \emph{{The
  perturbative CFT optical theorem and high-energy string scattering in AdS at
  one loop}}, \href{http://dx.doi.org/10.1007/JHEP04(2021)088}{\emph{JHEP} {\bf
  04} (2021) 088}, [\href{http://arxiv.org/abs/2012.01515}{{\tt 2012.01515}}].

\bibitem{Caron-Huot:2021kjy}
S.~Caron-Huot and Y.-Z. Li, \emph{{Helicity basis for three-dimensional
  conformal field theory}},
  \href{http://dx.doi.org/10.1007/JHEP06(2021)041}{\emph{JHEP} {\bf 06} (2021)
  041}, [\href{http://arxiv.org/abs/2102.08160}{{\tt 2102.08160}}].

\bibitem{Li:2021snj}
Y.-Z. Li, \emph{{Notes on flat-space limit of AdS/CFT}},
  \href{http://dx.doi.org/10.1007/JHEP09(2021)027}{\emph{JHEP} {\bf 09} (2021)
  027}, [\href{http://arxiv.org/abs/2106.04606}{{\tt 2106.04606}}].

\bibitem{Jain:2022ujj}
S.~Jain and A.~Mehta, \emph{{4D Flat-space scattering amplitude /$CFT_3$
  correlator correspondence revisited}},
  \href{http://arxiv.org/abs/2201.07248}{{\tt 2201.07248}}.

\bibitem{Kundu:2021qcx}
S.~Kundu, \emph{{Subleading Bounds on Chaos}},
  \href{http://arxiv.org/abs/2109.03826}{{\tt 2109.03826}}.

\bibitem{Kundu:2021mex}
S.~Kundu, \emph{{Extremal chaos}},
  \href{http://dx.doi.org/10.1007/JHEP01(2022)163}{\emph{JHEP} {\bf 01} (2022)
  163}, [\href{http://arxiv.org/abs/2109.08693}{{\tt 2109.08693}}].

\bibitem{Dolan:2003hv}
F.~Dolan and H.~Osborn, \emph{{Conformal partial waves and the operator product
  expansion}},
  \href{http://dx.doi.org/10.1016/j.nuclphysb.2003.11.016}{\emph{Nucl. Phys. B}
  {\bf 678} (2004) 491--507}, [\href{http://arxiv.org/abs/hep-th/0309180}{{\tt
  hep-th/0309180}}].

\end{thebibliography}\endgroup
\bibliographystyle{JHEP}

\end{document}